# FRIB Theory Users Group Report

## Joint ATLAS-HRIBF-NSCL-FRIB SuperUsers Meeting 18-20 August 2011

A. Baha Balantekin
Department of Physics, University of Wisconsin, Madison, WI 53706

Richard H. Cyburt, Filomena Nunes*, and Scott Pratt
National Superconducting Cyclotron Laboratory and Department of Physics,
Michigan State University, East Lansing, MI 48824

W. C. Haxton*
Department of Physics, University of California, Berkeley,
and Lawrence Berkeley National Laboratory, Berkeley, CA 94702

Witek Nazarewicz and Thomas Papenbrock
Department of Physics and Astronomy, University of Tennessee, Knoxville, TN 37996
and Oak Ridge National Laboratory, Oak Ridge, TN 37831

James Vary
Department of Physics and Astronomy, Iowa State University, Ames, IA 50011

*Corresponding authors

## Abstract

The FRIB (Facility for Rare Isotope Beams) Theory Users Group participated in the Joint ATLAS-HRIBF-NSCL-FRIB SuperUsers Meeting, hosted by Michigan State University August 18-20, 2011.  Prior to the meeting a survey of the FRIB Theory Users Group was conducted to assess the health of the low-energy nuclear theory community and to identify perceived areas of need, in anticipation of the new demands on theory that will accompany FRIB.  Meeting discussions focused on survey results and on possible responses.  These discussions are summarized here.

The Joint ATLAS-HRIBF-NSCL-FRIB SuperUsers Meeting, hosted by Michigan State University August 18-20, 2011, was attended by about 230 members of the low-energy nuclear physics community, including a significant contingent of theorists. Theory participation was organized by the FRIB Theory Users Group (FTUG).

The theory context for the meeting was set by two recent FTUG endeavors. The first was FTUG's initiation of a merger with the FRIB Users Organization (FUO), a move that was warmly welcomed by our experimental colleagues. The purpose of the merger is to better integrate experiment and theory as the FRIB program takes shape, so that theory can do its part to support FRIB intellectually, through working group participation, FRIB program planning, users meetings participation, etc.

The second endeavor was FTUG's recent "self study," which was conducted through a survey of the membership to identify important issues affecting FRIB theory support. The survey provided an accounting of the current distribution of research effort and also asked individuals, first from the perspective of their special subfields and second from a field-wide vantage, to identify areas needing additional theory support. The survey, described in more detail below, was the starting point for the SuperUsers Meeting theory discussion: given this assessment of where we are and where we hope to be at FRIB turn-on, what new steps should we be taking? While it will take time to formulate crisp answers to this question, this report summarizes the progress that was made during meeting discussions.

The list of theory participants in the theory sessions of the SuperUsers meeting can be found in Appendix I. Thirty FTUG members participated on site while an additional 16 joined the discussions remotely, through go-to-meeting. The agenda for the theory sessions is included as Appendix II. The plenary presentation on which this report is based and the plenary presentation on education (see the Education, Computation, and Personnel section) are available in their original forms at
http://meetings.nscl.msu.edu/superuser2011/program.htm.

### The Survey
The survey on FRIB theory addressed three issues:
- The fraction of research time individuals spend on various sub-areas of low energy nuclear physics, broken down into faculty or staff (PI), postdoc (PD), or student (S) contributions.
- Research not currently being pursued, but that would be undertaken if additional resources were available. Respondents were asked to specify the resources needed.
- The identification of sub-areas where additional theory support is critical.

Input was solicited from the community using the FTUG email list, supplement by targeted followed-up emails. In addition, the nuclear theory program officers from the DOE and the NSF provided a list of funded PIs in low energy nuclear physics, so that these PIs could be contacted. To identify sub-area needs, the field was first divided into four main areas, structure, reactions, nuclear astrophysics, and fundamental symmetries. The first three areas were then further subdivided into sub-areas:
- Structure: *ab initio* methods, configuration interaction methods, density functional theory and its extensions, algebraic methods and collective models, other models of nuclear structure;
- Reactions: *ab initio* reactions, direct reactions, fission/fusion, compound nucleus reactions, central collisions;
- Nuclear astrophysics: nucleosynthesis, supernova modeling, neutron stars.

A total of 170 responses were received from 68 senior researchers (university faculty or national laboratory staff), 40 postdocs and 62 students. Those respondents belonging to low-energy research groups but not reporting effort in the field were removed from the analysis, leaving 163 respondents (64 PIs, 40 postdocs and 59 students). Of these, 138 indicated that 100% of their effort is dedicated to low energy nuclear physics, while only eight reported less than 50%. The data from the survey are compiled in an excel file and are available on request.

Table 1 summarizes the effort per area, broken down into PI, PD and S, while Table 2 gives the percentage effort. The largest PI effort is under structure, accompanied by correspondingly larger numbers of postdocs and students. This is followed by reactions and astrophysics, with a rather meager fraction of overall PI effort going into fundamental symmetries.

Table 1: Effort per sub-area for principal investigators, postdocs and students. Also shown are the fraction of postdocs and students per PI per area.

|             | PI effort | PD effort | S effort | PD/PI | S/PI |
|-------------|-----------|-----------|----------|-------|------|
| Structure   | 26.8      | 18.9      | 28.5     | 0.7   | 1.1  |
| Reactions   | 18.1      | 6.5       | 9.8      | 0.4   | 0.5  |
| Astro       | 9.8       | 8.5       | 14.3     | 0.9   | 1.5  |
| Fund. Symm. | 2.1       | 3.5       | 3.4      | 1.7   | 1.7  |
| Total       | 56.7      | 37.4      | 56.0     |       |      |

The anomalously low values of PD/PI and S/PI for reactions were discussed during the meeting. Further investigation of the data revealed that, of the 18.1 PI effort reported under reactions, 12 PIs spend more that 50% of their time in reactions and of those, only five are university faculty, with roughly half at PhD granting institutions. The other seven PIs who spend more than 50% of their time on reactions are at national laboratories and most of these are either retired or nearing retirement. These factors likely account for the low PD/PI and S/PI ratios compared to other areas.

The very low numbers of PIs working on fundamental symmetries was another point of discussion at the SuperUsers meeting. We will return to this issue in the science section.

For completeness, in Table II we present the percentage distribution in the various areas for PIs, PD and S.

In the survey, each area was subdivided into sub-areas (Table 3). Within structure, the peak related to *ab initio* research is prominent, followed by a strong effort into energy density functional work and configuration interaction methods.

**Table 2: Percentage distribution per area**

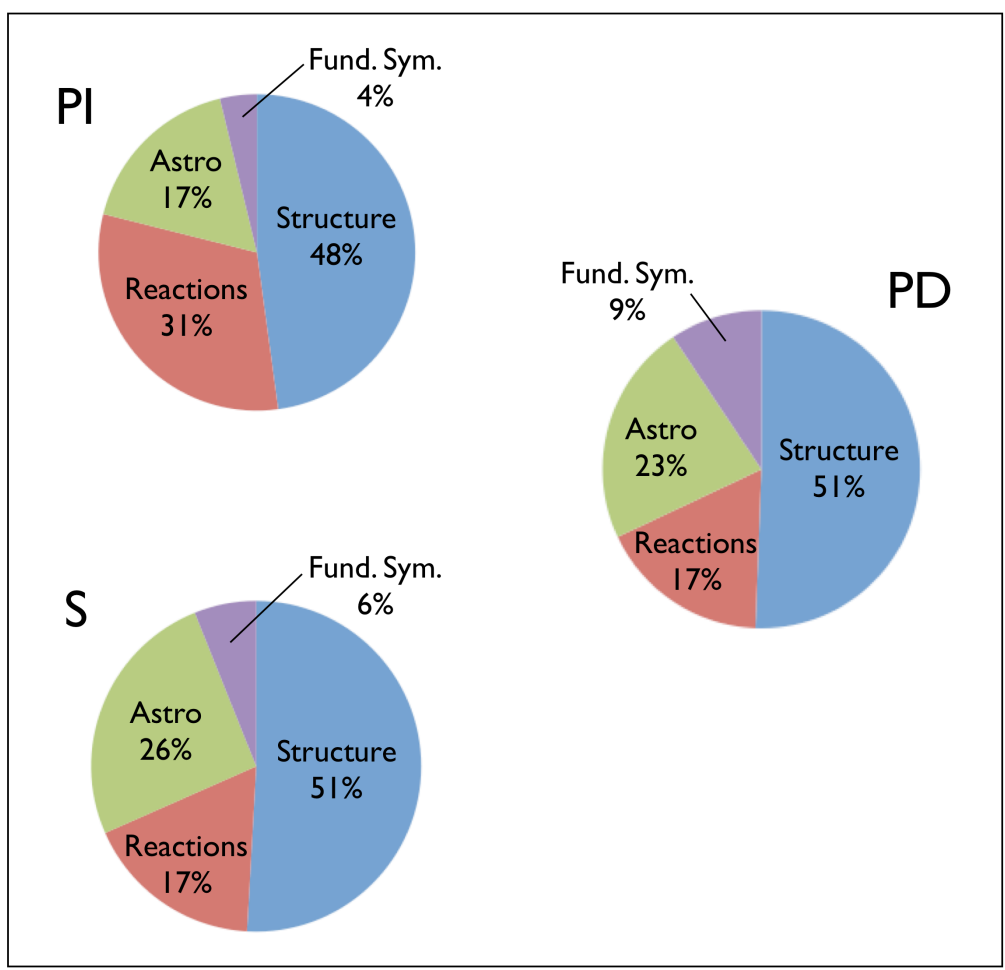

The distribution in the various sub-fields is much flatter. Although a significant effort is being invested in *ab initio* theory, a good fraction of the work is being carried out through students, in sharp contrast to the situation in direct reactions and fission/fusion. These variations appear to correlate in part with the research situations of the PIs (university versus national laboratory). Within reactions, we note that the effort in compound and central collisions is particularly low.

Within astrophysics, the neutron star effort appears to be quite strong. But some caution should be taken with these numbers because our mailing list sources were not sufficiently inclusive of astrophysics PIs, nor was the response from the included PIs as complete as desired. We will return to this point later.

In addition to the distribution of effort, the survey included two open-ended questions:
  i) Are there important developments that the group is not undertaking but would embark on if additional resources were made available? Specify which and what types of resources?
  ii) Of the sub-areas identify, are there specific ones in which you feel additional support is critical? Which?
Open-ended responses were provided by 45 of the 67 PIs.

**Table 3: Distribution of effort per sub-area**

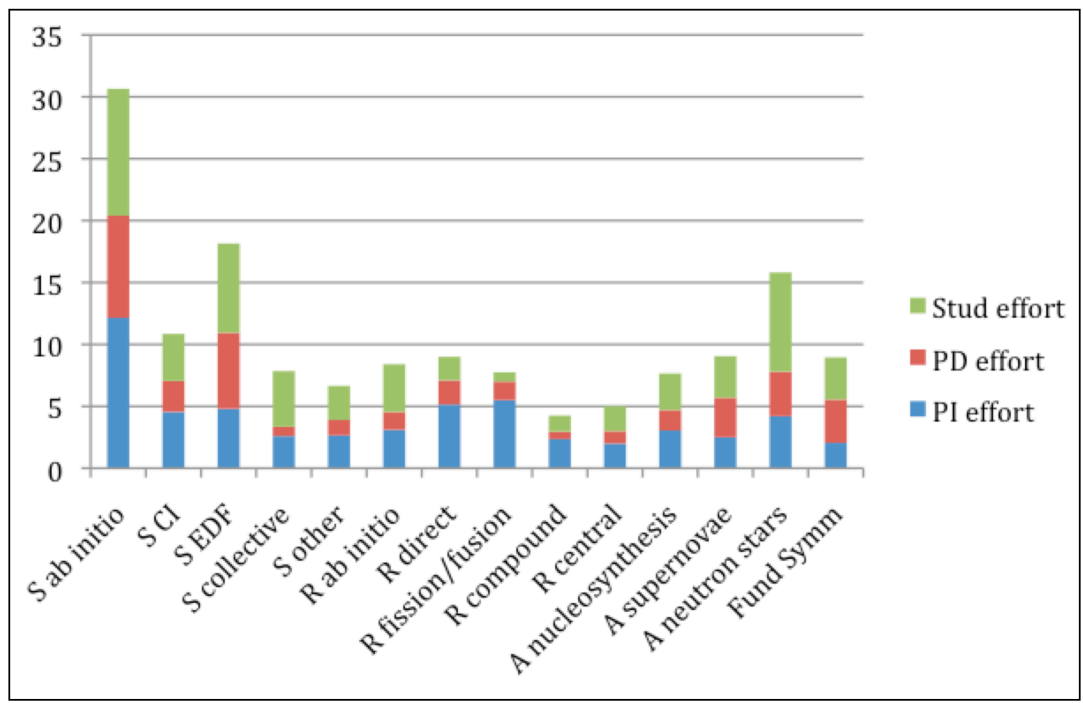

Responses to i) spanned a wide range of topics, with the common limitation being manpower: most referred explicitly to postdoctoral support, followed by lack of faculty colleagues and of graduate students. Also mentioned was the need for additional resources in computing.

Responses to ii) converged on two areas: reaction theory and fundamental symmetries. If responses are weighted by the number of PIs represented in a given survey, there are 25 PIs that identify a need in reaction theory, and 20 PIs that suggest fundamental symmetries as a critical area.

### SuperUser Meeting Discussions: Science

The survey results were the catalyst for presentations and discussions at the theory gathering during the SuperUsers meeting. Two parallel sessions were devoted to theory, with the first dedicated primarily to the survey results and to any "action items" the survey might suggest. The active participants included both the 30 attending theoreticians and the 16 "virtual" ones who participated through videoconferencing, commenting and asking questions from their remote sites. Presenters were asked to discuss the challenges facing each of the four areas mentioned above, taking into account survey results. Following each presentation, suggestions and discussions were solicited, with the goal of reaching consensus on major points. The progress made toward that goal is reported here.

The summary of goals for the **nuclear structure** community include:
- The development of more unified theoretical approaches to replace the current patchwork of methods and models; understanding the limitations of such approaches as they are pushed to the extremes of isospin, small binding energy, etc.
- Establish bridges so that FRIB data can impact a wider community: one would hope to develop a cascade of connections, e.g., lattice QCD → effective field theories → interactions → first-principles calculations → configuration interaction models → energy density functional theories and collective models.
- Attention to prediction: quantification of model uncertainties, validation and verification of nuclear structure tools.
- Understanding and delineating the connections between bound-state properties and reactions/scattering theory.
- Contributing to "national needs" problems such as energy production and national security.
- Formulating the intellectual arguments that connect nuclear structure studies to the goals of other subfields. Areas of opportunity include the role of reactions in astrophysics, the nuclear structure of neutron stars and supernovae, the analogy between the generic properties of nuclei and those of systems studied in atomic/condensed matter physics, and the physics of open quantum systems.

One observation made during the discussions is that the field is agile. The build-up in *ab initio* methods is recent and dramatic, illustrating what happens when new research directions coincide with new agency initiatives: numerically intensive research on *ab initio* methods found support in programs such as SciDAC and INCITE.

Important themes that emerged from the **reaction theory** presentation and discussion include

- The incorporation of microscopic input, including detailed structure and realistic effective interactions.
- Calibration of existing techniques with *ab initio* methods.
- Quantifying uncertainties.
- Assessing and then modernizing the tools of reaction theory.
- Learning from other fields, such as quantum chemistry and atomic physics.

Both the survey results and meeting discussions show that a strong community consensus exists that the reaction theory effort in the US should be strengthened. There are relatively few PIs in the field, and the number of postdocs and students per PI is anomalously low. Existing efforts are often single-PI or laboratory-based, making student recruitment more difficult. There is need for a strategy that will bring more young people into this area.

The summary for **nuclear astrophysics** included the following points:
- Nuclear structure is critical to astrophysics, impacting our understanding of nucleosynthesis and chemical evolution, the hydrostatic burning of stars, supernovae (of both types), a variety of explosive transient events, gamma-ray spectroscopy, neutron stars, and neutrino physics.
- Astrophysics needs include reaction rates, masses, and weak lifetimes; particle emission probabilities; and the properties of dense, neutron-rich nuclear matter.
- FRIB and similar facilities are the primary source of laboratory astrophysics data for the neutron-rich nuclei encountered in the r-process and in neutron star crusts, for the rp and $\nu$p processes, and for other explosive processes.
- Theory will have a special role in building the bridges between FRIB data and modelers and observers, so that nuclear physics becomes a participant in the astrophysics rather than just a convenient databank for others.

In meeting discussions, the last observation was connected to the survey, where we noted the low fraction of theoretical nuclear astrophysics PIs sampled (reflecting in part their level of participation in FTUG) and the disappointing response rate for those who were contacted. Somewhat in contrast to this situation, the FUO working group in (experimental) astrophysics is very vigorous. Consequently there is an opportunity for FTUG, especially as the merger with FUO progresses, to address this issue by reaching out to astrophysics theory PIs who can contribute to FRIB science.

The summary for **fundamental symmetries** included:
- Symmetry tests that exploit unusual nuclear properties to enhance interactions have exceptional potential, with nuclear electric dipole moment and beta decay asymmetry experiments being two examples.
- Success in this field would significantly broaden FRIB's scientific base, as this area has important intersections with fields such as atomic and particle physics.
- Among FRIB's four main legs, this may be the subfield that is least well organized experimentally. Bridges to those in nuclear physics who might pursue such experiments need to be strengthened.

- Theory leadership and ideas will be critical.  The subfield of fundamental symmetries emerged as the second case (the first being reactions) where the need for additional theory help is broadly recognized. This result came from a survey where almost all of the respondents were from outside this subfield.
- Based on the survey responses, only 2% of the PIs and 3% of the effort was in this area.  While the field is small, these numbers may also indicate that a significant fraction of the theorists working in this area are not yet connected to FRIB planning.
- Modest investments in the theory combined with efforts to engage key theorists could draw more attention to experimental opportunities for exploiting rare isotopes in novel symmetry tests.

As mentioned above in connection with nuclear astrophysics theory, FTUG may be able to help by engaging theorists not currently involved with FRIB – needed are theorists interested in the combination of nuclear structure and symmetry tests.  One opportunity FTUG will exploit is the recently approved summer 2013 INT program on nuclear structure and fundamental symmetries, which was proposed by FTUG members.  The weakness of the theory support for the FRIB fundamental symmetries program is of concern because experimental planning in this area may also be lagging.  Theory could be crucial in helping experimental colleagues identify opportunities for experiments that exploit rare isotopes, thereby inspiring a program.

## SuperUser Meeting Discussions: Education, Computation, Personnel

The sessions also included discussions of "infrastructure" issues such as student training, computing, and FRIB theory personnel programs.

**Education:**  The education discussion focused on a specific initiative, Training in Advanced Low-Energy Nuclear Theory (TALENT), focused on graduate students and young researchers (see http://nucleartalent.org).  This initiative is being pursued in both the US and Europe.  The proponents described the difficulty facing faculty wanting to teach nuclear theory, when university groups are often single-PI.  TALENT is a proposal to provide the needed advanced coursework through a program serving a variety of dispersed university groups.

TALENT will work with established researchers to develop a curriculum and to identify venues for the lectures.  It is envisioned that students and postdocs would attend the lectures, which would be delivered over a three-week period, while others including possibly senior researchers could participate through web broadcasts.  Materials would be archived on line, for future reference or repeat use.

Participating institutions and individuals would organize the schools for an audience consisting of both theorists and experimentalists.  The first courses would be held in 2012-13.  The possibility that students could get class credit from their home institutions is under discussion.

The TALENT discussion concluded with a suggestion that the TALENT initiative be sent to the full FTUG membership, for a possible endorsement vote.

**Computation for FRIB Theory:** Discussions focused on the rapid evolution in both the power and complexity of high performance computing (HPC), as illustrated by the DOE's program to develop extreme-scale computing in this decade. Nuclear physics has the capacity (the problems and the interest) to be a leader in scientific computing: we have reached a point where certain questions can be answered with HPC that neither analytic theory nor experiment can address. Some of these questions are of great significance to FRIB, including the modeling of explosive astrophysical environments, development of *ab initio* techniques for nuclear structure and reactions, and the building of nuclear energy density functionals.

The prospects for nuclear physics to become one of the fields that helps develop extreme-scale scientific computing (thereby advancing its use in other areas, including various national needs problems) is good, if the necessary resources can be found:
- The transition in HPC that is occurring – to machines that may have 1000 cores on one chip – resembles the transition our experimental colleagues made from basement tandems to JLab/RHIC/FRIB. The infrastructure requirements of leading edge HPC are growing rapidly. Theory does not have the capacity to manage this transition. It is important that HPC for nuclear physics be viewed as a important complement to both theory and experiment, and consequently treated as a third leg of the field.
- The sociological requirements for success are 1) a mechanism within the field to identify and develop important nuclear physics HPC applications; 2) the opportunity to develop the needed special relationships to applied mathematics and computer science, so that the broad-based "infrastructure" necessary to compete at the highest levels of scientific HPC are in place for nuclear physics; and 3) opportunities for young nuclear researchers to train simultaneously in their discipline and in HPC computing. As the applied mathematic and computer science expertise necessary for 2) and 3) resides largely in national laboratories, university-laboratory partnerships will be important.
- Nuclear physics will need adequate access to tier #1 facilities.
- Modest local facilities – clusters – will remain important for code developing, testing, pre/post processing, and student training.

**Personnel Issues:** FRIB is being established in part because the broader nuclear physics community recognizes that low-energy nuclear physics needs a flagship facility for community as well as scientific purposes. Community goals include student recruitment, new faculty positions, and establishing an intellectual center for a dispersed field.

Other flagship facilities have been coupled to personnel programs, e.g., the RIKEN/BNL and JLab junior-faculty positions. During the meeting there was a presentation on the RIKEN/BNL program, which is widely recognized as a success for heavy ion physics, and which therefore has been suggested as a possible model for FRIB.

As budgets are tight and may remain so for some time, it was also recognized that the FRIB program must be efficient and effective.  Goals identified at the meeting include
- Designing a program that produces a pipeline, beginning with graduate student recruitment and training, continuing in the postdoctoral years, then transitioning to faculty or other senior positions.
- A priority should be the placement of new faculty in strong research universities, where they can be effective in training their own students.
- The FRIB program should be national in focus, not regional.
- It should not be narrowly focused, but instead be designed to encompass as much talent as possible, in the manner of the RIKEN/BNL program.  FRIB has the advantage that its four scientific legs cover a great deal of nuclear physics.
- The program should be conducted as a community/agency partnership.  The building that is done must be sustainable, and thus requires long-term agency buy-in.   The researchers helped through faculty bridge positions will need grant support throughout their careers.

### FRIB Theory Users Group: Tasks Ahead
The FTUG Executive Committee is in the early stages of digesting the survey and assessing how theory can best support the FRIB program.  Indeed, this report can be considered an interim one describing initial ideas about how to respond to what we learned, as well as an invitation to the community (especially those who were unable to attend the SuperUsers Meeting) to comment.  Comments should be addressed to this report's corresponding authors.

The FTUG is anxious to complete the merger with FUO.  In addition to the administrative issues, associated tasks will include
- Updating our website, which will be reached through the FUO portal, and our mailing lists.
- Reviewing our bylaws for consistency with those of FUO.
- Merging the FTUG and FUO membership lists: anyone who belongs to FTUG will, once the merger is completed, be automatically a FUO member.
- Renewed efforts to broaden participation in FTUG, particularly in areas like fundamental symmetries and nuclear astrophysics, where the survey indicates additional outreach is needed.

FTUG governance will remain virtually unchanged.  Activities will continue to be coordinated by an Executive Committee elected by the FTUG membership.

The theorists who attended the SuperUsers meeting enjoyed the experience.  During the oral presentation of these remarks, it was proposed that FTUG would take part in FUO meetings every other year, with a significant parallel program in theory.   There was a strong desire expressed by FRIB management and others that FTUG consider participating every year.   Additional comments urging yearly participation were received after the close of the meeting.  The Executive Committee will consult with membership on this matter, but the very positive experience of theorists at the SuperUsers meeting does argue for the strongest possible theory-experiment coupling within the FUO.

Finally, the attending theorists were asked by experimental colleagues to suggest a possible "endorsement" that they could adopt in support of theory. Our suggestion was

"Theory is central to the scientific program of FRIB, for defining intellectual directions, exploring new opportunities, and establishing connections to other fields. We endorse efforts of the theory community to identify strategies and resources needed to move theory forward, develop new collaborations between theory and experiment, and educate the next generation of nuclear physicists."

The SuperUsers meeting participants unanimously endorsed this statement.

## Appendix I: List of theory participants

| | |
|---|---|
| Alexander Volya | avolya@fsu.edu |
| Anatoli Afanasfev | aa242@msstate.edu |
| Baha Balantekin | baha@physics.wisc.edu |
| Bao An Lee | Bao-An_Li@tamu-commerce.edu |
| Betty Tsang | tsang@nscl.msu.edu |
| Chuck Horowitz | horowit@indiana.edi |
| Dean Halderson | halderson@wmich.edu |
| Dick Furnstahl | furnstahl.1@osu.edu |
| Ed Brown | ebrown@pa.msu.edu |
| Emily Wang | emilywan@msu.edu |
| Erich Ormand | ormand@llnl.gov |
| Filomena Nunes | nunes@nscl.msu.edu |
| Fransesca Samarruca | fsammarr@uidaho.edu |
| Gaute Hagen | gautehag@gmail.com |
| Ian Thompson | i-thompson@llnl.gov |
| James Vary | jvary@iastate.edu |
| Janina Grineviciute | janina.grineviciute@wumich.edu |
| Jerry Draayer | draayer@sura.org |
| Jorge Piekarewicz | jpiekarewicz@fsu.edu |
| Luke Titus | titus@nscl.msu.edu |
| Mark Caprio | mcaprio@nd.edu |
| Mihai Horoi | horoi@phy.cmich.edu |
| Muslema Pervin | pervin@nscl.msu.edu |
| Naftali Auerbach | auerbach@post.tau.ac.il |
| Pawel Danielewicz | bogner@nscl.msu.edu |
| Rebecca Surman | surmanr@union.edu |
| Richard Cyburt | cyburt@nscl.msu.edu |
| Scott Bogner | bogner@nscl.msu.edu |
| Scott Pratt | prattsc@msu.edu |
| Stefano Gandolfi | stefano@lanl.gov |
| Thomas Papenbrock | tpapenbr@utk.edu |
| Vladimir Zelevinsky | zelevinsky@nscl.msu.edu |
| Wick Haxton | haxton@berkeley.edu |
| Witek Nazarewicz | witek@utk.edu |
| Wolfgang Bauer | bauer@pa.msu.edu |
| Yang Sun | sunyang@situ.edu.cn |

Remote participants
| | |
|---|---|
| Alex Brown | brown@nscl.msu.edu |
| Bruce Barrett | bbarrett@physics.arizona.edu |
| Joaquin Drut | joaquindrut@gmail.com |
| Charlotte Elster | elster@ohiou.edu |
| Jon Engel | engelj@physics.unc.edu |
| Jutta Escher | escher1@llnl.gov |
| Mesut Karakoc | |
| Sanjay Reddy | reddy@phys.washington.edu |
| Pierre Capel | capel@nscl.msu.edu |
| Kyle Wendt | |

**Appendix II: Agenda for FRIB theory sessions**

### Thursday Evening, August 18 (room BPS1420)

7:00pm-8:00pm   Theory dinner meeting

8:00pm-8:10pm   Introduction (Wick Haxton)

8:10pm-8:20pm   Merge with FRIB Users (Wick Haxton)

8:20pm-8:30pm   Presentation of results from the survey (Filomena Nunes)

8:30pm-9:00pm   Discussion (chair Baha Balantekin)

### Friday Afternoon, August 19 (room BPS1420)

1:30pm-2:25pm   Discussion-Survey (Baha Balantekin)

2:25pm-2:50pm   Education (Witek Nazarewicz)

2:50pm-3:15pm   Computation (James Vary)

3:15pm-3:35pm   BNL/RIKEN model (Scott Pratt)

3:35pm-3:50pm   Break

3:50pm-4:15pm   Structure (Thomas Papenbrock)

4:15pm-4:40pm   Reactions (Filomena Nunes)

4:40pm-5:05pm   Astrophysics (Richard Cyburt)

5:05pm-5:30pm   Fundamental (Wick Haxton)

5:35pm-6:05pm   Presentation to Tim Hallman